# Towards tunable graphene phononic crystals

*Yuefeng Yu[1]\*, Jan N. Kirchhof[1]\*, Aleksei Tsarapkin[2], Victor Deinhart[2,3], Oguzhan Yucel[1], Bianca Höfer[1], Katja Höflich[2] and Kirill I. Bolotin[1]*

\*Authors contributed equally to this work.

[1]Department of Physics, Freie Universität Berlin, Arnimallee 14, 14195 Berlin, Germany.

[2]Ferdinand-Braun-Institut gGmbH Leibniz-Institut für Höchstfrequenztechnik, Gustav-Kirchhoff-Str. 4, 12489 Berlin, Germany.

[3]Helmholtz-Zentrum Berlin für Materialien und Energie, Hahn-Meitner-Platz 1,

14109 Berlin, Germany.

**Abstract**

Phononic crystals (PnCs) are artificially patterned media exhibiting bands of allowed and forbidden zones for phonons. Many emerging applications of PnCs from solid-state simulators to quantum memories could benefit from the on-demand tunability of the phononic band structure. Here, we demonstrate the fabrication of suspended graphene PnCs in which the phononic band structure is controlled by mechanical tension applied electrostatically. We show signatures of a mechanically tunable phononic band gap. The experimental data supported by simulation suggest a phononic band gap at 28–33 MHz in equilibrium, which upshifts by 9 MHz under a mechanical tension of 3.1 N/m. This is an essential step towards tunable phononics paving the way for more experiments on phononic systems based on 2D materials.

**Introduction**

The periodic arrangement of atoms in solids gives rise to the band structure for electrons and with it to all the richness of solid-state physics. During the last decades, artificial periodic patterning of materials on a much larger length scale has been used to induce a band structure for phonons in so-called phononic crystals (PnCs)[1–4]. PnCs have been used as widely re-configurable analogues of solids to study phenomena ranging from metal-insulator transition to topological states and acoustic flatbands[5–7]. In addition, phononic band gaps have been employed to decouple localized vibrational modes from their environment resulting in ultra-coherent resonances suitable for storing quantum information[8–10].

There is one feature of traditional solid-state crystals that is so far elusive in PnCs: in-situ tunability. The band structure parameters of solids can be adjusted via e.g. electrostatic gating, application of magnetic fields, or mechanical straining. Such tunability allows a plethora of experiments ranging from the exploration of the phase transition to changing the band structure topology[11–16] or increasing the electron mobility in silicon[17–19] . In contrast, phonons from PnCs do not react to an electrical or magnetic field. In addition, many of the materials used to create PnCs, such as silicon or silicon nitride, are rigid and cannot be mechanically manipulated in-situ.

Graphene, as well as other two-dimensional materials, hold promise for tunable phononic crystals. From one perspective, graphene can be directly patterned down to a near-atomic length scale while supporting a variety of phononic lattices. In addition, graphene is a strong yet flexible material capable of withstanding tension as high as 42 N/m (~10% of strain) without breaking [20]. These properties, in combination, suggest mechanically-tunable graphene PnCs can be created[21]. Nevertheless, the challenges associated with creating large-area, uniform graphene membranes in combination with a challenging detection of high-order modes prevented experimental demonstration of tunable graphene PnCs so far.

Here, we demonstrate the design and characterization of a functional tunable graphene PnC. The device is made from a suspended large-scale (10 μm radius) few-layered graphene membrane, which is patterned with uniform honeycomb lattice holes. This step transforms it into a phononic crystal and gives rise to a phononic band structure, which ultimately determines any mechanical motion within the device. We use interferometric measurements to establish the band gap for out-of-plane acoustic phonons in the MHz range within the structure. We then show that additional tension up to >3 N/m can be induced using electrostatic gating. In response to that increase in tension, we observe signatures consistent with an upshift of the phononic band gap of 9 MHz (~29 %).

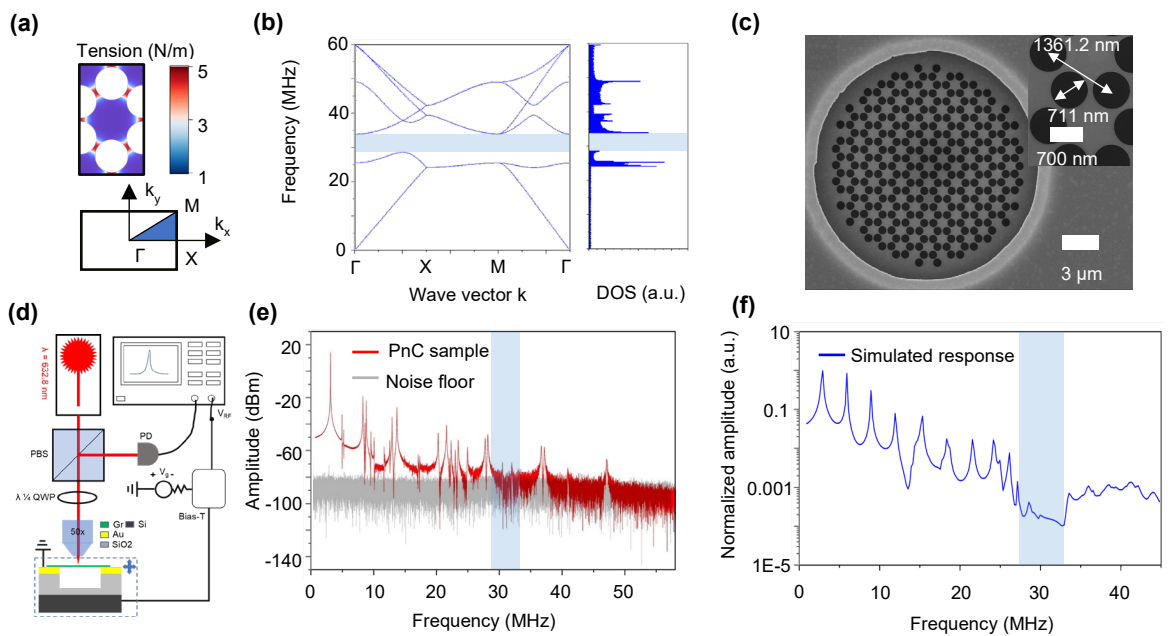

**Fig. 1: Device design and measurements.** *a) PnC unit cell and first Brillouin zone used for band structure calculations. b) Band structure for out-of-plane modes of the PnC obtained using an "infinite" model, along with the corresponding density of states. c) Fabricated device structure using the lattice parameters from a). d) Experimental setup used for interferometrically measuring phononic spectra. QWP is a quarter wave plate. e) Vibrational spectrum of the same device at zero applied gate voltage (pre-tension=0.44 N/m). The grey trace shows the noise floor of the setup (sample actuation off). The region of the phononic band gap is shown in blue. In the band gap region, we find a strongly suppressed response, which falls below the noise floor of the experiment. f) Simulated vibration spectrum of the corresponding devices.*

**Structure design and fabrication**

Our design of a mechanically-tunable PnC requires a suspended graphene membrane so that the tension can be controlled by applying a voltage. This puts the natural limitation on device size as it is hard to create free-standing graphene with radii larger than 10 μm[22]. In order to improve the uniformity and to make the sample less sensitive to surface contaminations, we use a trilayer graphene. To create the phononic crystal structure, we pattern a regular honeycomb lattice of holes (diameter *d*) in the suspended graphene membrane. The unit cell and the corresponding first Brillouin zone of this lattice are shown in Fig. 1a. The position and size of the phononic band gap in the structure depend on lattice parameter *a*, the ratio *d/a*, and the initial tension $\sigma$ in the membrane. We chose *d/a* = 0.522 which maximizes the band gap size. For choosing the lattice parameter, we need to find a compromise. Ideally, we want to make *a* as small as possible, to fit a large number of unit cells into the finite suspended area. However, small lattice constants result in a high frequency range for the band gap, which will make the detection challenging, in terms of suitable detectors and electronics. Therefore, we use a lattice parameter of *a* = 1.36 μm which corresponds to the PnC with approximately 7 unit-cells across in the armchair direction and 13 in the zigzag direction. Our fabrication protocol allows us reproducible pattern PnCs down to a lattice parameter $a$ ~100 nm. For smaller parameter values, the bridge separating the holes becomes thinner than 10 nm, and samples tend to break at this point. The final parameter of our PnC – the tension in the sample – is beyond our control. An annealing step before pattering helps to reduce the tension as possible and cleans the device from fabrication-related residues. We calculate the phononic band structure of the PnC with these parameters and a build-in tension of 0.44 N/m (obtained from experiments shown below) by matching the vibrational frequencies to a simulation in COMSOL (Fig. 1b). For this phononic lattice, we find a band gap between 28 and 34 MHz for out-of-plane modes. We note that only out-of-plane acoustic phonon modes are considered here, as they are most relevant in phononic crystals made from 2D materials because they are easy to excite and detect.

To fabricate this structure, we use Electron Beam lithography (EBL) and metallization on a $SiO_2$/Si wafer to define a hard mask for a subsequent wet-etching step (details in SI). After the etching, we have a ~1 μm deep pit in the substrates on which we transfer an exfoliated trilayer graphene flake using the polymer (Polydimethylsiloxane, PDMS) stamp dry transfer[23–26]. This leaves us with a suspended graphene membrane, which is electrically contacted via the gold layer. We chose thin multilayer exfoliated graphene to reduce wrinkling in the device, for band gap hunting and sample surviving. Finally, the phononic lattice discussed above shown in Fig. 1a is cut in the graphene using focus-ion beam (FIB) (Fig. 1c) [27]. In designing the device, our goal was to obtain high-uniformity device with predictable boundary conditions.

**Phononic crystal measurements**

The PnCs are measured inside a homebuilt cryostat with base pressure better than $1.3 \times 10^{-5}$ mbar (Fig. 1d). To drive the device, we use an electrical drive (an AC+DC voltage applied between the graphene and Si back-gate, Fig. 1d) for its high efficiency and because it induces less heating or photodoping than e.g. optothermal drive. By adjusting the DC component of the applied voltage, we can tune the applied electrostatic pressure on the phononic crystal and thus the tension in the suspended area. For detection, we use a 632.8-nm HeNe laser that is focused on the PnC. The suspended membrane and the back gate form a Fabry-Perot interferometer and the reflected signal is thus modulated by the oscillation of the membrane[28]. We detect this signal in an avalanche photodetector (PD in Fig. 1c). To increase signal-to-noise ratio, we insert a 5-mm aperture before the photodetector.

The measurements of PnCs require resolving high-order vibrational modes which in general have very low vibrational amplitude. Therefore, multiple approaches to optimize the sensitivity of our detection scheme are needed. We adjust the cavity length in the etching step to maximize the interferometric signal (see SI), we also use a polarized laser beam in combination with a polarization beam splitter and a quarter wave plate (QWP) that allows us to guide almost 100% of the reflected light into the photodetector. Finally, we implement a pinhole, that cuts out unwanted reflected light, and thereby reduces the laser noise in our detection scheme (details in SI). Overall, we obtain the displacement sensitivity better than 0.5 pm/$\sqrt{Hz}$ (see SI).

The measured vibrational spectrum of our PnC is shown in Fig. 1e. We observe the fundamental mode at 2.97 MHz, and ~35 subsequent modes above the noise level. We track the fundamental modes vs. applied gate voltage and extract the built-in tension $T_0$ as well as the mass density $\rho_{2D}$ of the device from fitting the capacitive softening regime. We find $T_0 = 0.44\ N/m$ and $\rho_{2D} = 5.2 \times 10^{-5} kg/m^2$, respectively. These parameters are used as inputs for simulations in Fig. 1b and 1f, see detail in SI. We prefer to use tension rather than strain as our key parameter since the former ultimately determines the frequency of our resonator ($f \sim \sqrt{T}$) while the latter depends on Young's modulus of the patterned device that is hard to measure directly. The extracted value for mass density is about 23 times higher compared to what is expected for pristine graphene, likely due to the impurities always present on the surface of 2D materials.

While the find a densely populated "forest" of vibrational modes at frequencies above the fundamental mode, we see a region of strong suppression between 28 and 33 MHz (blue in Fig. 1e). This region is exactly where the band gap is expected from our band structure calculation. To better understand this spectrum, we carry out a "finite" simulation accounting for the finite size of the PnC and considering boundary conditions of the device. We can directly compare our experimental vibrational spectra (Fig. 1e) to the outcome of the simulation (Fig. 1f). We see a similar phononic band gap in both graphs, as well as similar peaks on both sides of the band gap, likely to the peaks in the phononic density of states associated with flat bands at both sides of the band gap (Fig. 1b). We note that the simulation does not match the measurements perfectly. This is likely due to unavoidable stress non-uniformity in our devices caused by nonlinearity of cut drum-head and variation in boundary conditions. While frequency of modes from circular-membrane system are strongly affected by these effects, the position of the phononic band gap is relatively insensitive to them. It is also noteworthy that both the simulation and experiments show modes inside the band gap. These are localized states related to the finite size non-patterned edges of our system (see detail in SI). Finally, it should be noted that the measurements are limited by the size of the laser spot size preventing us from resolving most of the modes with frequency much higher than the band gap. When we compare our data to measurements of non-patterned devices (see SI), we find only a few (~5-8) lowest frequency modes compared to $N > 35$ in the patterned PnC sample. Also, we do not find a region of suppressed displacement in the reference samples (See SI). Both points highlight the effects of the phononic patterning in our devices.

**Tuning phononic crystals**

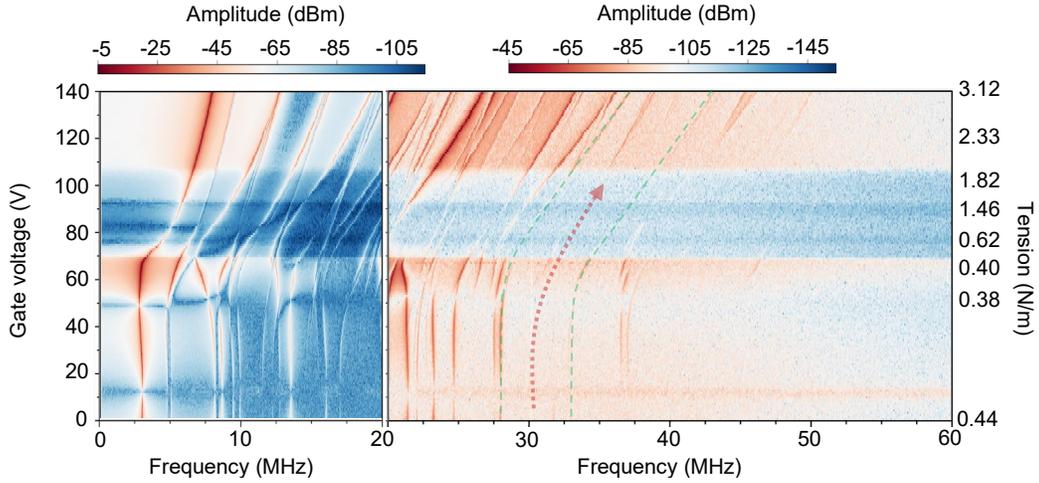

**Fig. 2: Tension-dependence of the PnC.** *Vibrational spectrum of a phononic crystal device vs. applied gate voltage controlling the tension. The regions around the fundamental mode as well as the entire spectrum are shown separately. Dashed green lines denote edges of the phononic band gap.*

Our next goal is to explore changes in vibrational spectra with induced tension and show in-situ tunability of our phononic system. As we vary the DC component of the gate voltage, we observe smooth evolution of the vibrational spectrum with the fundamental mode downshifting from 2.96 MHz to 2.74 MHz then upshifting to 7.92 MHz (Fig. 2). The initial downshift is due to the so-called capacitive softening[29–32], while upshift is caused by the increase of tension. The higher order modes show a variety in different tuning behavior, which we attribute to a combination of photothermal backaction effects[33–35] and different coupling to the electrostatic field for different modes[36–38]. Whilst there are interesting interactions taking place, this is not the scope of the work. We assume that the application of DC voltage produces tension $T = \frac{f^2}{f_0^2}T_0$, where $f_0$ and $f$ are the values of the fundamental mode frequency before and after the application of gate voltage. We find that we generate up to $T-T_0$=2.68 N/m external tension at our highest $V_g$ = 140 V.

We now analyze the evolution of the phononic band gap region, from 28 to 33 MHz at zero $V_g$. Experimentally, we find that the region upshifts together with other modes. Using an effective spring constant extracted from the fundamental modes, we can simulate the evolution of the phononic band gap (green line in Fig. 2). We find that that region upshifts by 9 MHz at our highest tension. The region preserves a low vibrational amplitude. At the same time, we observe the appearance of new vibrational peaks inside it. This behavior also seen in simulations likely corresponds to new defect modes activated by the gradual violation of the translational symmetry and stronger drive power at higher voltages. Overall, the data in Fig. 2 suggest that the phononic band gap in the graphene phononic crystals is mechanically tunable by 9 MHz (29 %).

**Conclusions**

In conclusion, we developed an approach to fabricating suspended gate-tunable graphene phononic crystals and observed signatures of phononic band gap variation in these devices. These results suggest several interesting pathways for further work. First, it is instructive to consider the factors limiting the quality of our graphene PnCs. One of them is the device size limiting the number of unit cells inside the PnC (currently 57). Better growth and transfer techniques may help to increase this number. Another limitation is non-uniformity inside the device coming primarily from uneven edges of holes pattern, unavoidable wrinkling in graphene and gold contacts. We anticipate that it may be possible to produce more uniform devices by employing "additive" manufacturing strategies where additional material is deposited onto graphene to create periodic PnCs. Finally, the quality of our spectra is limited by the diffraction-limited size of our read-out probe making it challenging to access higher-lying vibration modes. Near-field techniques such as Scanning Near-field Optical Microscope (SNOM) may be employed to break this barrier[39,40]. In the future, it will be especially interesting to consider potential applications of tunable phononic systems. These include, for instance, tension-tunable in-gap mechanically shielded defect modes as well as exploring mechanical analogues to phase transition in condensed matter systems. Finally, it will be interesting to explore the role of the inherent non-linearity of graphene on phononic spectra. Our devices are especially suitable for this task since the patterning effectively softens the membrane resulting in large oscillation amplitudes and therefore strong non-linear effects.

**Acknowledgments**

This work was supported by Deutsche Forschungsgemeinschaft (DFG, German Research Foundation, project-ID 449506295 and TRR 227 B08, 328545488), EU COST Action CA 19140 (FIT4NANO) and CSC 202006150013. The Ga ion beam patterning was performed in the Corelab Correlative Microscopy and Spectroscopy at Helmholtz-Zentrum Berlin.

**Reference**

1. Maldovan, M. Sound and heat revolutions in phononics. *Nature* **503**, 209–217 (2013).
2. Zen, N., Puurtinen, T. A., Isotalo, T. J., Chaudhuri, S. & Maasilta, I. J. Engineering thermal conductance using a two-dimensional phononic crystal. *Nat. Commun.* **5**, 3435 (2014).
3. Pennec, Y., Vasseur, J. O., Djafari-Rouhani, B., Dobrzyński, L. & Deymier, P. A. Two-dimensional phononic crystals: Examples and applications. *Surf. Sci. Rep.* **65**, 229–291 (2010).
4. Cha, J. & Daraio, C. Electrical tuning of elastic wave propagation in nanomechanical lattices at MHz frequencies. *Nat. Nanotechnol.* **13**, 1016–1020 (2018).
5. Kirchhof, J.N., Bolotin, K.I. Mechanically-tunable bandgap closing in 2D graphene phononic crystals. npj 2D Mater Appl 7, 10 (2023).

6.  Ren, H. *et al.* Topological phonon transport in an optomechanical system. *Nat. Commun.* **13**, 1–7 (2022).

7.  Zhang, Q. *et al.* Gigahertz topological valley Hall effect in nanoelectromechanical phononic crystals. *Nat. Electron.* **5**, 157–163 (2022).

8.  Beccari, A. *et al.* Strained crystalline nanomechanical resonators with quality factors above 10 billion. *Nat. Phys.* **18**, 436–441 (2022).

9.  Rossi, M., Mason, D., Chen, J., Tsaturyan, Y. & Schliesser, A. Measurement-based quantum control of mechanical motion. *Nature* **563**, 53–58 (2018).

10. Zivari, A., Stockill, R., Fiaschi, N. & Gröblacher, S. Non-classical mechanical states guided in a phononic waveguide. *Nat. Phys.* **18**, 789–793 (2022).

11. Takashina, K., Ono, Y., Fujiwara, A., Takahashi, Y. & Hirayama, Y. Valley polarization in Si(100) at zero magnetic field. *Phys. Rev. Lett.* **96**, 2–5 (2006).

12. Tse, J. S. *et al.* Pressure-induced changes on the electronic structure and electron topology in the direct FCC → SH transformation of silicon. *J. Phys. Chem. C* **118**, 1161–1166 (2014).

13. Noborisaka, J., Nishiguchi, K. & Fujiwara, A. Electric tuning of direct-indirect optical transitions in silicon. *Sci. Rep.* **4**, 1–6 (2014).

14. Xu, E. G. B. X. Z. Two-dimensional silica: Structural, mechanical properties, and strain-induced band gap tuning . *J. Appl. Phys.* **119**, 014301 (2016).

15. Huang, B., Monsma, D. J. & Appelbaum, I. Experimental realization of a silicon spin field-effect transistor. *Appl. Phys. Lett.* **91**, (2007).

16. Tan, K. Y. *et al.* Transport Spectroscopy of single phosphorus donors in a silicon nanoscale transistor. *Nano Lett.* **10**, 11–15 (2010).

17. Tiwari, S., Fischetti, M. V., Mooney, P. M. & Welser, J. J. Hole mobility improvement in silicon-on-insulator and bulk silicon transistors using local strain. *Tech. Dig. - Int. Electron*

*Devices Meet. IEDM* **62**, 939–941 (1997).

18. Gleskova, H. *et al.* Field-effect mobility of amorphous silicon thin-film transistors under strain. *J. Non. Cryst. Solids* **338–340**, 732–735 (2004).

19. Kravchenko, S., Simonian, D., Sarachik, M. & Kent, A. Effect of a tilted magnetic field on the anomalous conducting phase in high-mobility Si MOSFET's. *Phys. Rev. B - Condens. Matter Mater. Phys.* **58**, 3553–3556 (1998).

20. Changgu Lee, Xiaoding Wei, Jeffrey W. Kysar & James Hon. Measurement of the Elastic Properties and Intrinsic Strength of Monolayer Graphene. *Science.* **321**, 382–385 (2008).

21. Kirchhof, J. N. *et al.* Tunable Graphene Phononic Crystal. *Nano Lett.* **21**, 2174–2182 (2021).

22. Steeneken, P. G., Dolleman, R. J., Davidovikj, D., Alijani, F. & Van Der Zant, H. S. J. Dynamics of 2D material membranes. *2D Mater.* **8**, (2021).

23. Jayasena, B. & Melkote, S. N. An Investigation of PDMS Stamp Assisted Mechanical Exfoliation of Large Area Graphene. *Procedia Manuf.* **1**, 840–853 (2015).

24. Seo, J. *et al.* Direct Graphene Transfer and Its Application to Transfer Printing Using Mechanically Controlled , Large Area Graphene / Copper Freestanding Layer. *Adv. FUnc. Mater.* **28**, 1707102 (2018).

25. Raimond, J. M., Brune, M., Computation, Q., Martini, F. De & Monroe, C. Electric Field Effect in Atomically Thin Carbon Films. **306**, 666–670 (2004).

26. Suk, J. W. *et al.* Transfer of CVD-Grown Monolayer Graphene onto Arbitrary Substrates. 6916–6924 (2011).

27. Deinhart, V. *et al.* The patterning toolbox FIB-o-mat: Exploiting the full potential of focused helium ions for nanofabrication. *Beilstein J. Nanotechnol.* **12**, 304–318 (2021).

28. J. Scott Bunch, Arend M. van der Zande, Scott S. Verbridge, Ian W. Frank, David M. Tanenbaum, Jeevak M. Parpia, Harold G. Craighead, P. L. M. Electromechanical Resonators from Graphene Sheets. *Science.* **315**, 490–493 (2007).

29. Song, X. *et al.* Stamp transferred suspended graphene mechanical resonators for radio frequency electrical readout. *Nano Lett.* **12**, 198–202 (2012).

30. Mei, T., Lee, J., Xu, Y. & Feng, P. X. L. Frequency tuning of graphene nanoelectromechanical resonators via electrostatic gating. *Micromachines* **9**, (2018).

31. Weber, P., Güttinger, J., Tsioutsios, I., Chang, D. E. & Bachtold, A. Coupling graphene mechanical resonators to superconducting microwave cavities. *Nano Lett.* **14**, 2854–2860 (2014).

32. Eichler, A. *et al.* Nonlinear damping in mechanical resonators made from carbon nanotubes and graphene. *Nat. Nanotechnol.* **6**, 339–342 (2011).

33. Barton, R. A. *et al.* Photothermal Self-Oscillation and Laser Cooling of Graphene Optomechanical Systems. *Nano Lett.* **12**, 4681–4686 (2012).

34. Favero, I. & Karrai, K. Optomechanics of deformable optical cavities. *Nat. Photonics* **3**, 201–205 (2009).

35. MacCabe, G. S. *et al.* Nano-acoustic resonator with ultralong phonon lifetime. *Science.* **370**, 840–843 (2020).

36. Wang, Z. *et al.* Resolving and Tuning Mechanical Anisotropy in Black Phosphorus via Nanomechanical Multimode Resonance Spectromicroscopy. *Nano Lett.* **16**, 5394–5400 (2016).

37. De Alba, R. *et al.* Tunable phonon-cavity coupling in graphene membranes. *Nat. Nanotechnol.* **11**, 741–746 (2016).

38. Singh, V. *et al.* Probing thermal expansion of graphene and modal dispersion at low-temperature using graphene nanoelectromechanical systems resonators. *Nanotechnology* **21**, (2010).

39. Prezelj, J., Nikonov, A. & Emri, I. Using sound in the very near field of vibrating plates for determination of their mechanical properties. *Appl. Acoust.* **186**, 108486 (2022).

40. Chen, X. *et al.* Modern Scattering-Type Scanning Near-Field Optical Microscopy for Advanced Material Research. *Adv. Mater.* **31**, 1804774 (2019).

# Supplementary information

## I. Sample fabrication

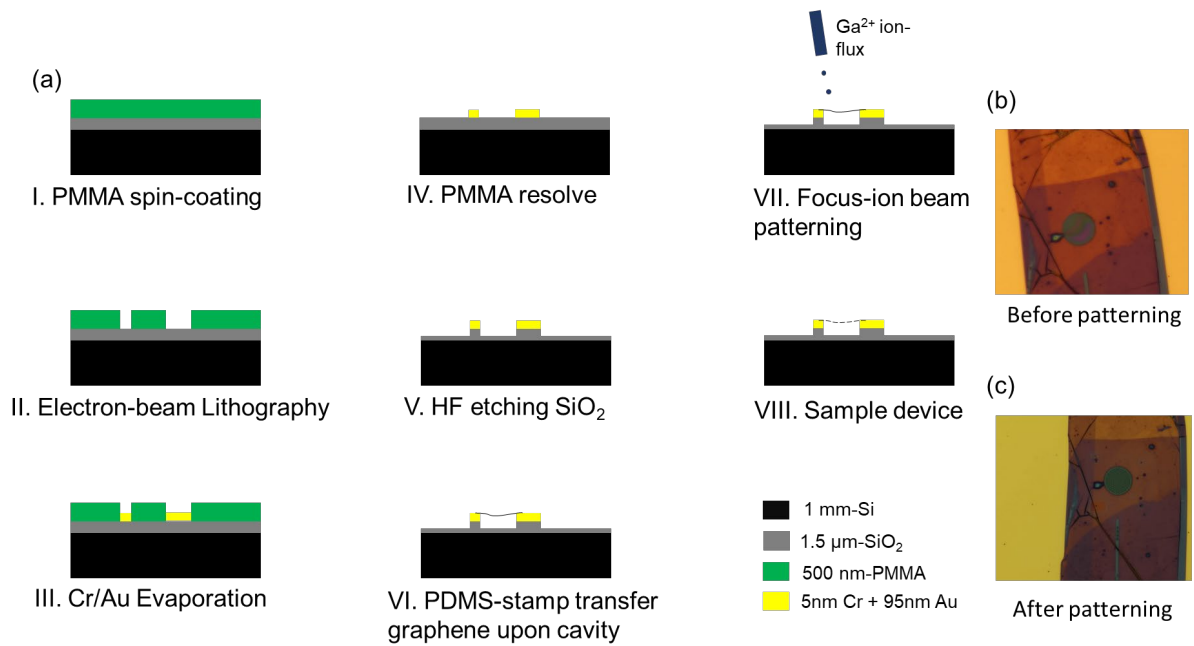

**Fig. S1: Procedure of sample fabrication** *a) fabrication of substrate chip and device. b) SEM image before b) and after c) FIB patterning*

Device are fabricated starting with Si/SiO$_2$ wafer pieces (0.5 cm*0.5 cm width/length; thickness of SiO$_2$ is 1.5um). The procedure of substrate fabrication is shown in Fig. S1. In step III, 5nm Cr/95nm Au is evaporated. In step V, about 1060nm of the SiO$_2$ layer is etched leaving the rest of the oxide as an insulating layer. Such a depth enables strong interferometric signal ($g_0 = (2N + 1) \cdot \frac{\lambda}{4}$, N is positive integer and rest of parameters are explained in section II, III) and a large deflection of the membrane center and therefore large strain. In step VI, graphene is transferred onto the cavity. At this point, large variations in membrane's thickness are observed (red regions of the membrane seen in Fig. S1b). These variations disappear after FIB patterning of the membrane (step VIII).

## II. Sensitivity calibration under IF bandwidth of 10 Hz

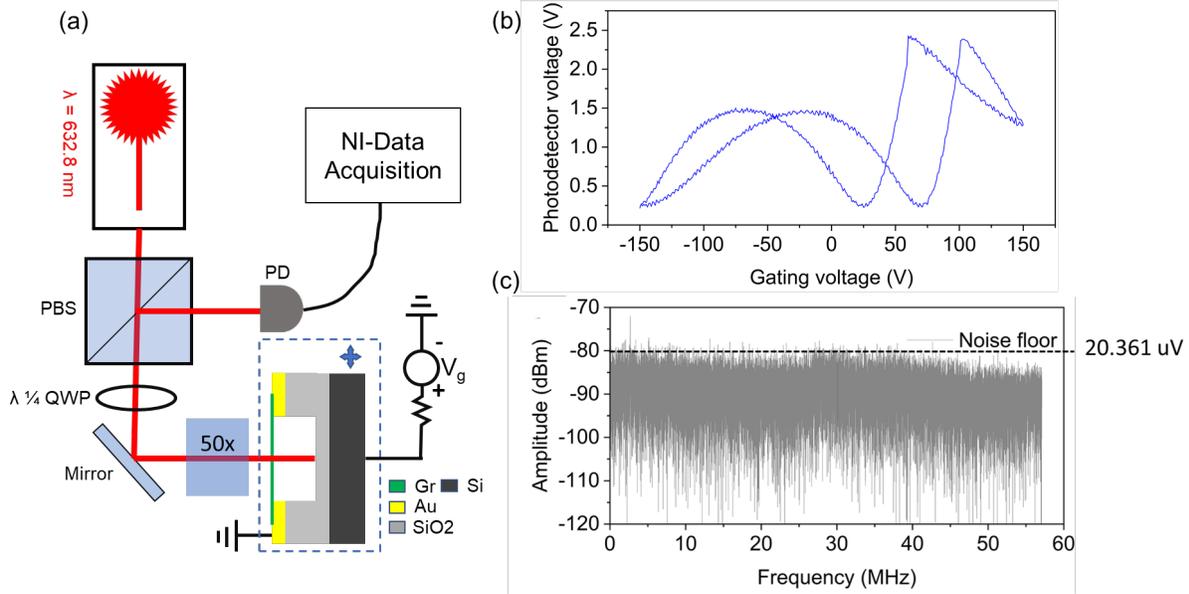

**Fig. S2: Characterization of measurement sensitivity** *a) Experimental setup b) Voltage on a photodetector vs. gating voltage. Lowest reflection is seen when the signal on photodetector is $V_{min}=0.235$ V, the highest signal is $V_{max}=2.414$ V. c) Noise floor spectrum measured by disabling electrical actuation.*

Our goal is to quantitatively determine the amplitude of mechanical vibration of the membrane. To do this, we need to convert the signal recorded by the photodetector (PD) in the units of Volt into displacements in the unit of meter [1,2]. In Fig. S2b, we show the PD signal ($V_{PD}$) vs. the gate voltage. The periodic variation of the amplitude corresponds to the membrane moving across the standing-wave pattern inside the cavity. We can analytically estimate the signal on the photodetector assuming it is proportional to the light intensity reflected from the cavity containing graphene:

$$I_{tot} \propto \text{const}(P_{in}) + \cos(\varphi_{gr}(z) - \varphi_0) \quad (1)$$

Here $\varphi_{gr} = \frac{2\pi}{\lambda} \cdot 2z$ is the phase acquired by the light wave on the travel across the cavity, and $\varphi_0$ is a phase shift dependent on the SiO$_2$ thickness, which can be defined as reference phase shift, $\varphi_0 = 0$. $P_{in}$ is the incident laser power, determining the constant term in eq.S1. We see that the light intensity goes from a minimum to a maximum when the membrane is moved by the distance $z = \frac{\lambda}{4}$. Experimentally, we see that a minimum and a maximum of intensity correspond to the photodiode voltages $V_{min} = 0.235\,V$ to $V_{max} = 2.414\,V$. Assuming that for most of our experiments the membrane is located between the maximum and a minimum, we can therefore determine the amplitude of a vibration of the membrane $z_{ac}$ to the amplitude of the signal of on the photodetector $\widetilde{V_{ac}}$ using linear interpolation:

$$z_{ac} = \frac{V_{max} - V_{min}}{\widetilde{V_{ac}} \cdot \lambda/4} \quad (2)$$

Finally, we can obtain the amplitude of noise in our system. We disable electrical actuation by grounding the sample and record a spectral noise amplitude (Fig. S2c). From the noise floor of the measurement, around -80 dBm, we estimate the sensitivity of the measurement into a 10Hz IF bandwidth:

$$s = \frac{\widetilde{V_{ac}}}{\frac{V_{top} - V_{bot}}{\lambda/4} * \sqrt{IF}} < 0.5 \text{ pm}/\sqrt{Hz} \quad (3)$$

### III. Built-in tension and membrane density

The goal of this section is to estimate the built-in tension as well the density of the membrane from experimental data. In general, the resonance frequency of a suspended circular membrane in a field effect transistor geometry, with voltage $V$ applied between the membrane is the gate below is evaluated in Ref. [3]:

$$f^2(V) = \frac{1}{4m\pi^2}\left[\frac{2\pi Eh\epsilon}{1-v^2} + \frac{(1-v)\pi\varepsilon_0^2 r^2}{8(1+v)Eh\epsilon^2 g_0^4}V^4 - \frac{0.542\varepsilon_0\pi r^2}{2g_0^3}V^2\right] \qquad (4)$$

where m is the effective mass, $\epsilon$ is built-in strain, $E$ is Young's modulus of a suspended membrane after patterning, h = $10^{-9}$ m is its effective thickness, $v = 0.15$ is the Poisson ratio, r = $9.9*10^{-6}$ is radius of the suspended region, $g_0$ = $1.206*10^{-6}$ m is the effective separation between membrane's center and the bottom gate. The effective separation is defined as $g_0 + \frac{t_{SiO_2}}{\varepsilon_{SiO_2}}$, where g is the membrane/SiO$_2$ distance and $t_{SiO_2}$ – thickness of SiO$_2$. Treating $E\epsilon$ and $m$ as free parameters we fit the experimental data in the region $V < 66$ V.

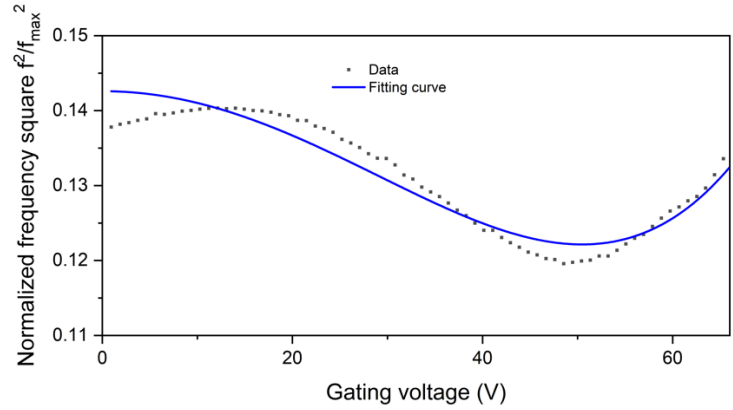

**Fig.S3: fundamental-vibrational-mode frequency response from gating voltage**

From the fit, we obtain $E\epsilon = 436\ MPa$, effective mass $m = 1.06 * 10^{-14}\ kg$. Area of patterned membrane $A = \pi r^2 - N\pi r_h^2$, where $r_h$ = is the radium of single hole in pattern, $N$ = 270 is the total number of cut holes in the pattern. Converting, we obtain $\rho_{2D} = \frac{m}{A} = 5.2 \times 10^{-5} kg/m^2$, built-in tension is $T_{built-in} = E\epsilon h = 0.44\ N/m$.

We note that our fitting only works in the regime of small gate voltage, the region of so-called capacitive softening [3–5]. At large gate voltage, we see that the fit does not describe the data well. We hypothesize that this is related to the buckling transition in our patterned membrane.

## IV. Midgap modes inside the phononic bandgap

Our experiments suggest that there are modes localized withing the region of a phononic bandgap at high strain. In Fig. S4, we demonstrate that such modes appear in a simulation accounting for boundary condition and finite device size.

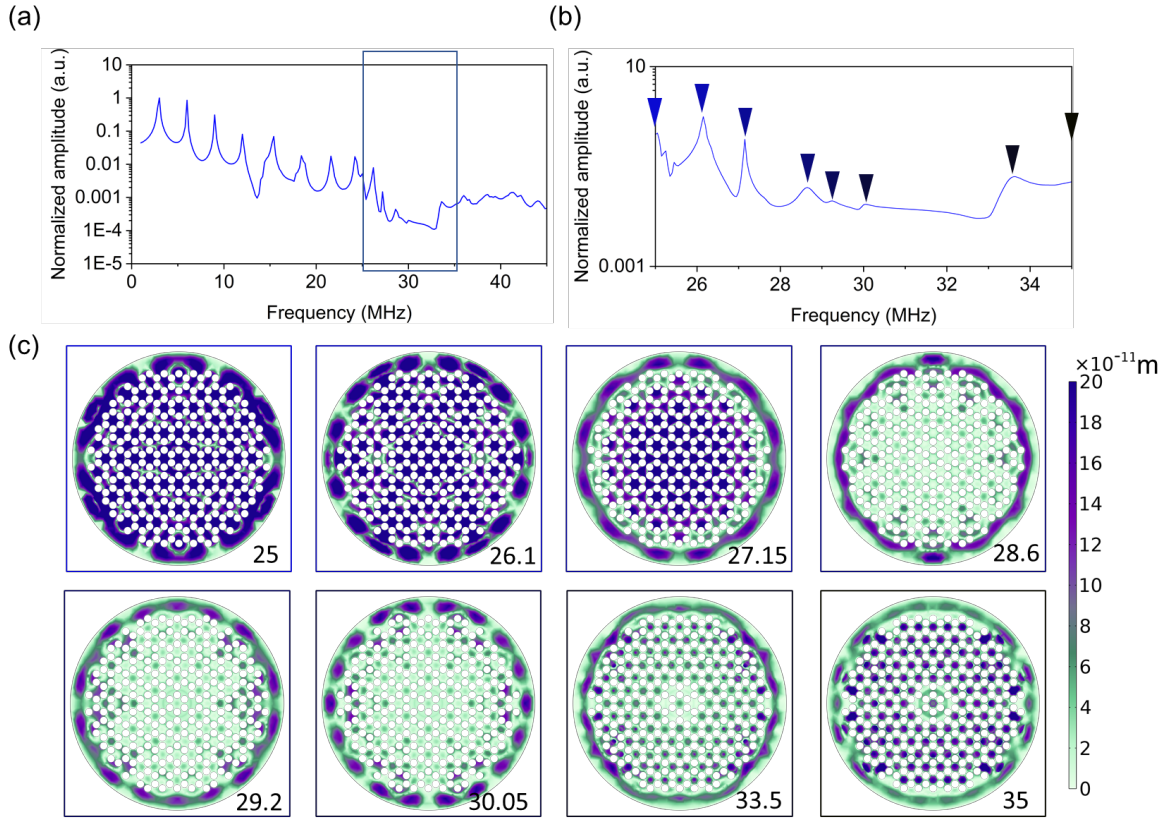

**Fig. S4: Midgap modes in a finite-size phononic crystal** *(a) Resonance spectrum, same as in Fig.1f. (b) Zoom-in of the bandgap region 25–35MHz showing a number of modes inside the bandgap region (c) Spatial distribution of the amplitude of several of the modes visible in (b) (The frequency of each mode in MHz is shown in the bottom right corner). All of the modes are localized*

*at the edge of the membrane and do not appear in "infinite" simulations neglecting finite device size.*

### V. Comparing patterned and unpatterned devices

To confirm that the observed phononic bandgap originate from a pattern of holes cut in our device, we fabricated, measured and simulated a graphene device without patterning. We only detect first 4 modes in such a device and no signature of a bandgap (Fig. S5a). Likewise, COMSOL simulation of phononic spectrum of the unpatterned device exhibit multiple vibrational modes with roughly constant mode density without any signature of a bandgap region (Fig. S5b).

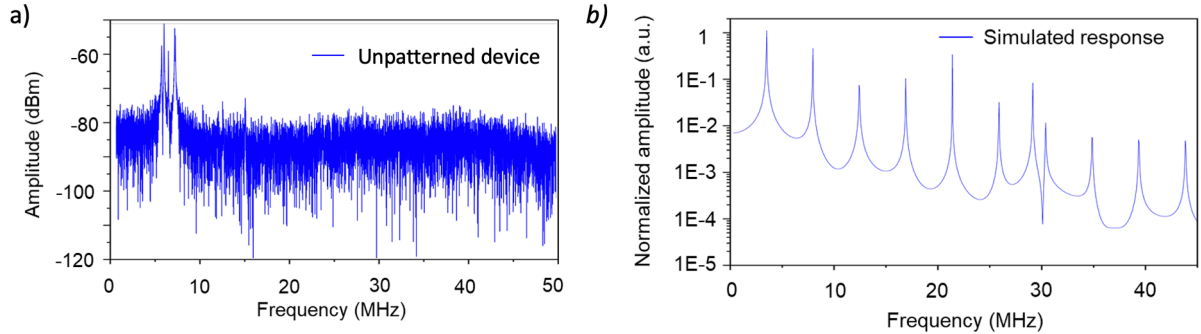

**Fig. S5: Vibration of graphene drum head w/o pattern.** *(a) Resonance response spectrum of an unpatterned graphene drum under strong drive of 8 dBm. (b) Simulated response of (a).*

### Reference

1. Callera Aguila, M. A. *et al.* Fabry-Perot interferometric calibration of van der Waals material-based nanomechanical resonators. *Nanoscale Adv.* **4**, 502–509 (2022).
2. Esmenda, J. C. *et al.* Optoelectrical Nanomechanical Resonators Made from Multilayered

Two-Dimensional Materials. *ACS Appl. Nano Mater.* **5**, 8875–8882 (2022).

3. Chen, C. Graphene NanoElectroMechanical Resonators and Oscillators. *Thesis Columbia* (2013).

4. Weber, P., Güttinger, J., Tsioutsios, I., Chang, D. E. & Bachtold, A. Coupling graphene mechanical resonators to superconducting microwave cavities. *Nano Lett.* **14**, 2854–2860 (2014).

5. Šiškins, M. *et al.* Nanomechanical probing and strain tuning of the Curie temperature in suspended Cr2Ge2Te6-based heterostructures. *npj 2D Mater. Appl.* **6**, 1–8 (2022).